\documentclass[oneside,english]{amsart}
\usepackage[T1]{fontenc}
\usepackage[latin9]{inputenc}
\usepackage{bm}
\usepackage{amstext}
\usepackage{amsthm}
\usepackage{graphicx}
\usepackage{esint}

\makeatletter

\providecommand{\tabularnewline}{\\}

\numberwithin{equation}{section}
\numberwithin{figure}{section}
  \theoremstyle{remark}
  \newtheorem*{rem*}{\protect\remarkname}

\makeatother

\usepackage{babel}
  \providecommand{\remarkname}{Remark}

\begin{document}

\title{Effective diffusion in the region between two surfaces}

\author{Carlos Valero Valdes\\
Departamento de Matematicas\\
Universidad de Guanajuato\\
 Guanajuato, Gto, Mexico}

\date{1 November 2015}

\maketitle
\global\long\def\CC{\mathbb{C}}

\global\long\def\RR{\mathbb{R}}

\global\long\def\zz{\bm{z}}

\global\long\def\xx{\bm{x}}

\global\long\def\yy{\bm{y}}

\global\long\def\nn{\bm{n}}

\global\long\def\ee{\bm{e}}

\global\long\def\kk{\bm{k}}

\global\long\def\ff{\bm{f}}

\global\long\def\jj{\bm{j}}

\global\long\def\vv{\bm{v}}

\global\long\def\dd{\bm{d}}

\global\long\def\pp{\bm{p}}

\global\long\def\YY{\bm{Y}}

\global\long\def\XX{\bm{X}}

\global\long\def\ZZ{\bm{Z}}

\global\long\def\JJ{\bm{J}}

\global\long\def\NN{\bm{N}}

\global\long\def\grad{\bm{\nabla}}

\global\long\def\dv{\bm{\nabla}\cdot}

\global\long\def\tangent{T}

\global\long\def\SO{\hbox{SO}}

\global\long\def\GL{\hbox{GL}}

\global\long\def\tr{\hbox{tr}}

\global\long\def\der#1#2{\frac{\partial#1}{\partial#2}}

\global\long\def\dder#1#2{\frac{\partial^{2}#1}{\partial#2^{2}}}

\global\long\def\dero#1{\frac{\partial}{\partial#1}}

\global\long\def\DD{\mathcal{D}}

\global\long\def\d#1#2{\frac{d#1}{d#2}}

\global\long\def\and{\hbox{\,\, and\,\,\,}}

\global\long\def\where{\hbox{\,\, where\,\,\,}}

\global\long\def\SS{\mathcal{S}}

\global\long\def\EE{\mathcal{E}}

\global\long\def\QQ{\mathcal{Q}}

\begin{abstract}
The purpose of this paper is to provide a formula for the effective
diffusion operator $\DD$ obtained by projecting the 3-dimensional
diffusion equation onto a 2-dimensional plane, assuming reflective
boundary conditions at two surfaces in 3-dimensional space. The formula
provided corresponds to the case of finite transversal stabilization
rate in contrast to the infinite transversal stabilization rate formula
provided in \cite{va:edfibrebundles}. 
\end{abstract}

\section{Introduction}

The problem of understanding spatially constrained diffusion plays
an important role in diverse areas such as biology, chemistry and
nano-technology. Solving the diffusion equation for general constraining
geometries is a very difficult task. One way to deal with this obstacle
is to reduce the degrees of freedom of the problem by considering
only the main directions of transport. More concretely, consider the
diffusion equation
\[
\der Pt(x,t)=D_{0}\Delta P(x,t)\where x=(x_{1},\ldots,x_{n})
\]
with reflective boundary conditions in the border of the region of
interest. By integrating this equation along adequate transversal
directions to the main directions of transport, we obtain (in an approximate
manner) a diffusion equation of the form 
\[
\der pt(y,t)=\grad\cdot\left(\DD(y)\grad p(y,t\right))\where y=(y_{1},\ldots,y_{m})\and m<n.
\]
The $m\times m$ matrix $\DD(y)$ in the above formula is known as
\emph{the effective diffusion matrix}, $p$ is the \emph{effective
density function}, and $\grad\cdot$ and $\grad$ are the divergence
and gradient operators in an adequate metric in the $y$-variables
(see \cite{va:edfibrebundles}). The estimates of $\DD$ fall into
two categories.
\begin{enumerate}
\item \emph{Infinite transversal diffusion rate}. In this case it is assumed
the the density function $P$ stabilizes infinitely fast in directions
transversal to the main directions of transport. In \cite{va:edfibrebundles}
we have given a very general formula for $\DD$ for arbitrary values
of $m$ and $n$ with $m<n$. This formula contains as special cases
the results in \cite{kn:ogawa} and \cite{va:fj3dcurves} given for
$n=3$ and $m=1$.
\item \emph{Finite transversal diffusion rate}. In this case it is assumed
the the density function $P$ stabilizes in finite time in directions
transversal to the main directions of transport. This case has been
studied extensively for $n=2$ and $m=1$ in articles like \cite{kn:entropybarrierzwanzig},
\cite{kn:bradley}, \cite{kn:reguerarubi}, \cite{kn:kp-diffusion-projection}.
\end{enumerate}
In this article we provide a formula for the effective diffusion matrix
in the finite transversal rate case, obtained by projecting the diffusion
equation in $x,y,z$ variables onto the $x,y$ variables. We assume
that reflective boundary conditions hold on the region bounded by
surfaces of the form $z=z_{1}(x,y)$ and $z=z_{2}(x,y)$, where $z_{1}(x,y)<z_{2}(x,y)$.
Our results are summarized in Table \ref{tab:Results}.

\begin{table}
\begin{tabular}{c|l}
Effective diffusion matrix $\DD$ & Width function and fields\tabularnewline
\hline 
 & \tabularnewline
$\DD=D_{0}\left(\begin{array}{cc}
\omega & -\omega\mu\sin(\psi)\\
-\rho\sin(\psi) & \cos^{2}(\psi)+\mu\rho\sin^{2}(\psi)
\end{array}\right)$ & $w=z_{2}-z_{1}$\tabularnewline
 & \tabularnewline
$\psi=\arcsin\left(\frac{\grad z_{1}\cdot\grad^{\bot}z_{2}}{\sqrt{1+|\grad z_{1}|^{2}}\sqrt{1+|\grad z_{2}|^{2}}}\right)$ & $\grad w,\grad^{\bot}w$\tabularnewline
 & \tabularnewline
$m_{i}=\frac{\grad z_{i}\cdot\grad w}{(\grad z_{i}\cdot\grad^{\bot}w)\sin(\psi)+|\grad w|\cos(\psi)}$ & $\grad=\left(\dero x,\dero y\right)$\tabularnewline
 & $\grad^{\bot}=\left(-\dero y,\dero x\right)$\tabularnewline
$\rho+i\omega=\left(\frac{1}{m_{2}-m_{1}}\right)\log\left(\frac{1+im_{2}}{1+im_{1}}\right)$ & \tabularnewline
 & \tabularnewline
$\mu=\frac{m_{1}+m_{2}}{2}$ & \tabularnewline
 & \tabularnewline
\multicolumn{1}{c}{} & \tabularnewline
\end{tabular}\caption{\label{tab:Results}Effective diffusion matrix $\protect\DD$ for
surfaces $z=z_{1}(x,y)$ and $z=z_{2}(x,y)$. The matrix is computed
in the basis formed by the fields $\protect\grad w,\protect\grad^{\bot}w$.}
\end{table}

The outline of the article is as follows.
\begin{itemize}
\item In section 2 we recall the effective continuity equation and the effective
diffusion matrix in the\emph{ infinite transversal rate case}, and
explain the technique used in the next section to derive the formula
for the effective diffusion matrix in the \emph{finite transversal
rate case}.
\item In section 3 we compute a formula for the effective diffusion matrix
$\DD$ in the finite transversal rate case for two planes in 3-dimensional
space, and construct an effective diffusion ellipsoid $E_{\DD}$ which
will help us to understand the geometric and physical properties of
$\DD$.
\item In section 4 we use the results in the the previous sections to compute
a formula for the effective diffusion matrix (in the finite transversal
rate case) for two surfaces of the form $z=z_{1}(x,y)$ and $z=z_{2}(x,y)$. 
\item In section 5 we apply our formula for $\DD$ to an specific examples.
In particular, we recover the results in \cite{kn:di-projection-diffusion}
(for channels in the plane) as a special case of our more general
formula.
\end{itemize}

\section{The effective diffusion equation}

Consider the region in $3$-dimensional space given by the set of
points $(x,y,z)$ that satisfy 
\begin{equation}
z_{1}(x,y)\leq z\leq z_{2}(x,y),\label{eq:Region E defining equations}
\end{equation}
where $z_{1}=z_{1}(x,y)\and z_{2}=z_{2}(x,y)$ are scalar functions.
We are interested in the \emph{continuity equation}
\begin{equation}
\der Pt+D_{0}\dv\JJ=0,\label{eq:Continuity Equation}
\end{equation}
where $P=P(x,y,z,t)$ is the \emph{concentration density} and $\JJ=\JJ(x,y,z,t)$
is the \emph{density flux} vector field, with reflective boundary
conditions
\begin{equation}
\JJ(x,y,z_{i}(x,y),t)\cdot\nn_{i}(x,y)=0\label{eq:Reflective Conditions}
\end{equation}
where the unit normal $\nn_{i}$ to the surface $z=z_{i}(x,y)$ is
given by 
\[
\nn_{i}=\frac{(-\grad z_{i},1)}{\sqrt{1+|\grad z_{i}|^{2}}}.
\]
The \emph{effective concentration density} is given by 
\[
p(x,y,t)=\int_{z_{1}(x,y)}^{z_{2}(x,y)}P(x,y,z,t)dz
\]
and the \emph{effective density flux }by\emph{ }
\begin{equation}
\jj(x,y,t)=(j_{1}(x,y,t),j_{2}(x,y,t)),\label{eq:Effective Flux}
\end{equation}
where for $\JJ=(J_{1},J_{2},J_{3})$ we have that
\[
j_{i}(x,y,t)=\int_{z_{1}(x,y)}^{z_{2}(x,y)}J_{i}(x,y,z)dz.
\]
If the continuity equation \ref{eq:Continuity Equation} and the reflective
boundary conditions \ref{eq:Reflective Conditions} hold, we have
the \emph{effective continuity equation (see \cite{va:edfibrebundles})}
\begin{equation}
\der pt+\dv\jj=0.\label{eq:Effective Continuity Equation}
\end{equation}
We will assume that \emph{Fick's law}, i.e 
\begin{equation}
\JJ=-D_{0}\grad P,\label{eq:Fick's Law}
\end{equation}
for a constant scalar value $D_{0}$, so that the continuity equation
becomes the \emph{diffusion equation}
\begin{equation}
\der Pt=D_{0}\Delta P.\label{eq:Diffusion Equation}
\end{equation}

\subsection*{Infinitely transversal diffusion rate case }

In this case $P$ must be constant along the $z$-variable, i.e
\[
P=P(x,y,t),
\]
and the effective continuity equation \ref{eq:Effective Continuity Equation}
becomes (see \cite{va:edfibrebundles}) 
\[
\der pt(x,y,t)=\dv\left(w(x,y)\grad\left(\frac{p(x,y,t)}{w(x,y)}\right)\right),
\]
where the gradient and divergence operators in the above formula are
given by
\[
\grad p=\left(\der px,\der py\right)\and\grad\cdot\jj=\der{j_{1}}x+\der{j_{2}}y,
\]
which are the gradient and divergence operators in flat space. In
\cite{va:edfibrebundles} we showed that this equation can be written
as a diffusion equation
\begin{equation}
\der pt(x,y,t)=\dv\left(\grad p(x,y,t)\right),\label{eq:Effective Diffusion Equation (Infinite Case)}
\end{equation}
if the divergence and gradient operators used are the ones associated
to the metric tensor
\begin{equation}
g_{w}(x,y)=w(x,y)\left(\begin{array}{cc}
1 & 0\\
0 & 1
\end{array}\right),\label{eq:Width Metric Tensor}
\end{equation}
i.e
\begin{eqnarray*}
\grad p & = & \frac{1}{w}\left(\der px,\der py\right),\\
\dv\jj & = & \frac{1}{w}\left(\dero x(wj_{1})+\dero y(wj_{2})\right).
\end{eqnarray*}
A consequence of this is that we can study the effective diffusion
equation \ref{eq:Effective Diffusion Equation (Infinite Case)} by
simulating random walks with steps constructed using geodesic segments
of the metric $g_{w}$.

\subsection*{Finetely transversal diffusion rate case}

We are looking for a matrix-valued function $\DD=\DD(x,y)$ such that
the projection of the diffusion process in the $x,y$-plane is modeled
by an equation of the form
\[
\der pt(x,y,t)=\grad\cdot(\DD(x,y)\grad p(x,y,t)),
\]
where the gradient and divergence operators $\grad\cdot$ and $\grad$
are the ones associated with the metric tensor \ref{eq:Width Metric Tensor}. 

\begin{figure}
\includegraphics{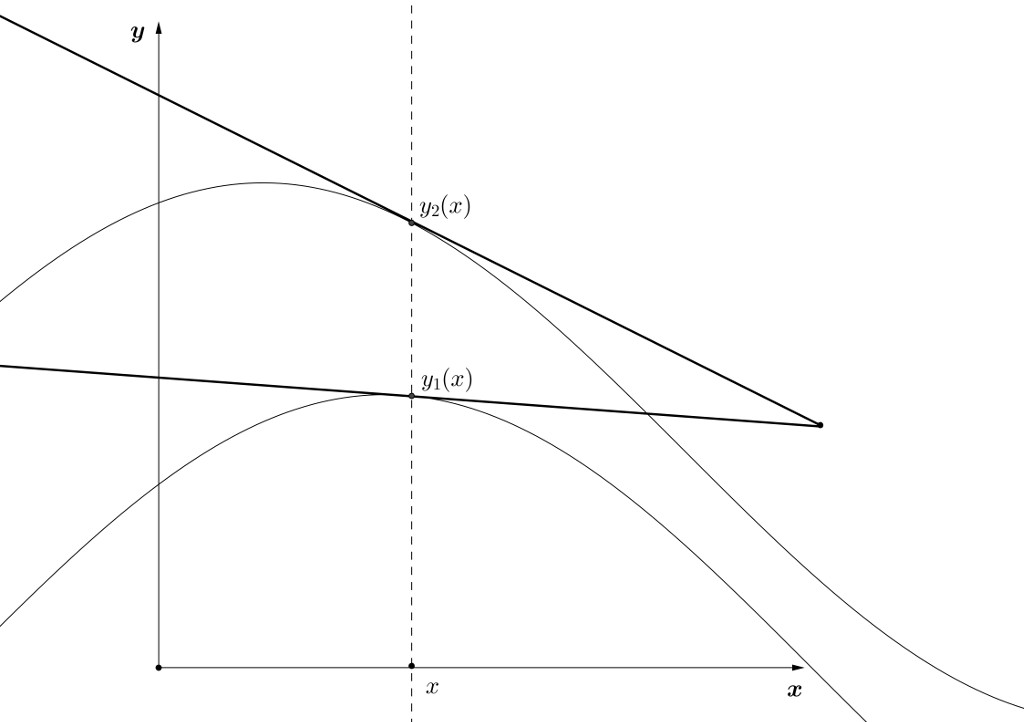}\caption{\label{fig:Channel2D Approximation}Approximation of a channel by
an angular sector}
\end{figure}

To compute $\DD$ we consider harmonic functions $Q=Q(x,y,z)$ (i.e
stable solutions of the diffusion equation) satisfying reflective
boundary conditions in the region of interest. If Fick's law holds,
i.e 
\[
\JJ_{Q}=-D_{0}\left(\der Qx,\der Qy,\der Qz\right),
\]
the effective density and flux are given by
\begin{eqnarray*}
q & = & \int_{z_{1}}^{z_{2}}Qdz,\\
\jj_{Q} & = & -D_{0}\left(\int_{z_{1}}^{z_{2}}\der Qxdz,\dero y\int_{z_{1}}^{z_{2}}\der Qydz\right).
\end{eqnarray*}
We can then compute $\DD$ from the formula (see \cite{kn:aproximations})
\[
\jj_{Q}=-w\DD\left(\grad\left(\frac{q}{w}\right)\right),
\]
to obtain 
\begin{equation}
\DD\left(\grad\left(\frac{q}{w}\right)\right)=\frac{D_{0}}{w}\left(\int_{z_{1}}^{z_{2}}\der Qxdz,\int_{z_{1}}^{z_{2}}\der Qydz\right).\label{eq:Effective Diffusion Formula (IDR)}
\end{equation}
In general it is not always possible to find explicit formulas for
harmonic functions $Q$ with the required boundary conditions. In
such cases we approximate the region of interest with a simpler one
in which we can find such functions (see Figure \ref{fig:Channel2D Approximation})

\section{Effective diffusion matrix for two planes}

\begin{figure}
\includegraphics{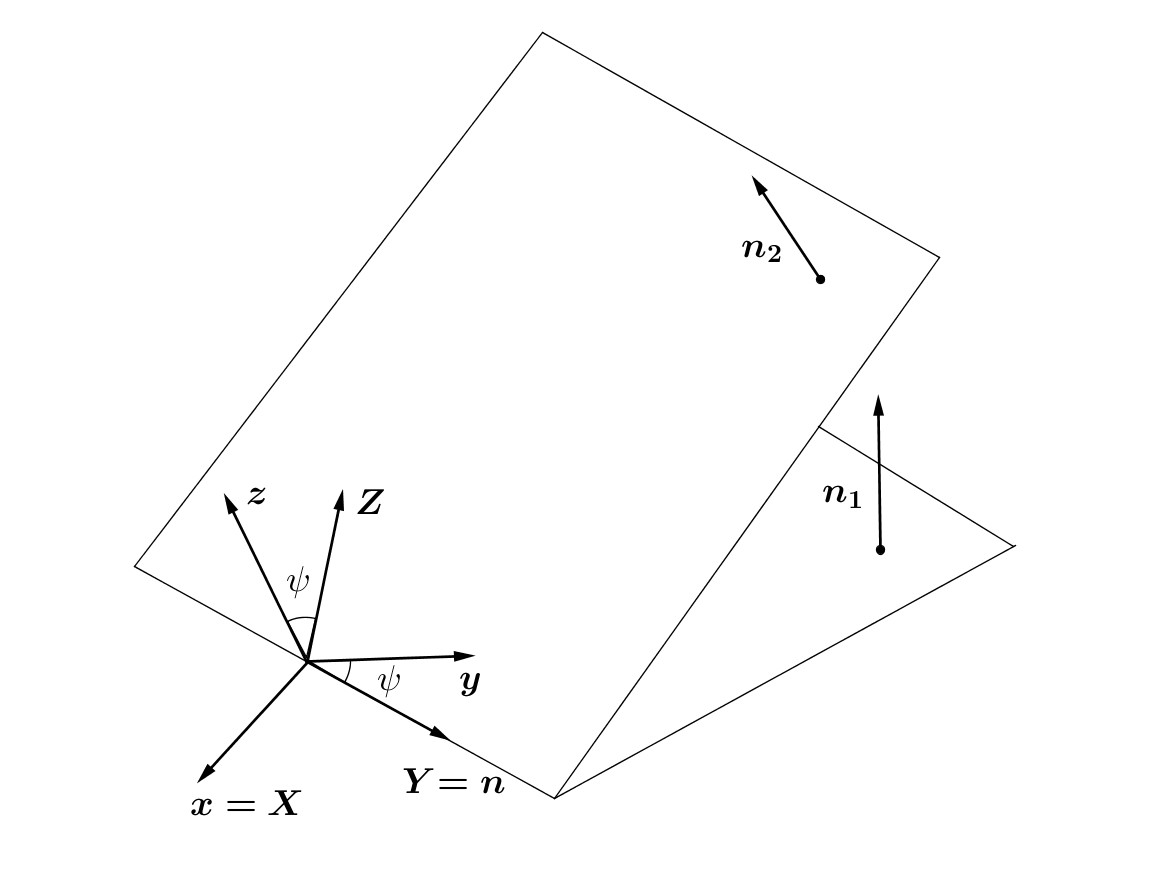}\caption{\label{fig:Projection frame}Intersecting planes and ortho-normal
basis $\protect\xx,\protect\yy,\protect\zz$.}
\end{figure}

\begin{figure}
\includegraphics[scale=0.7]{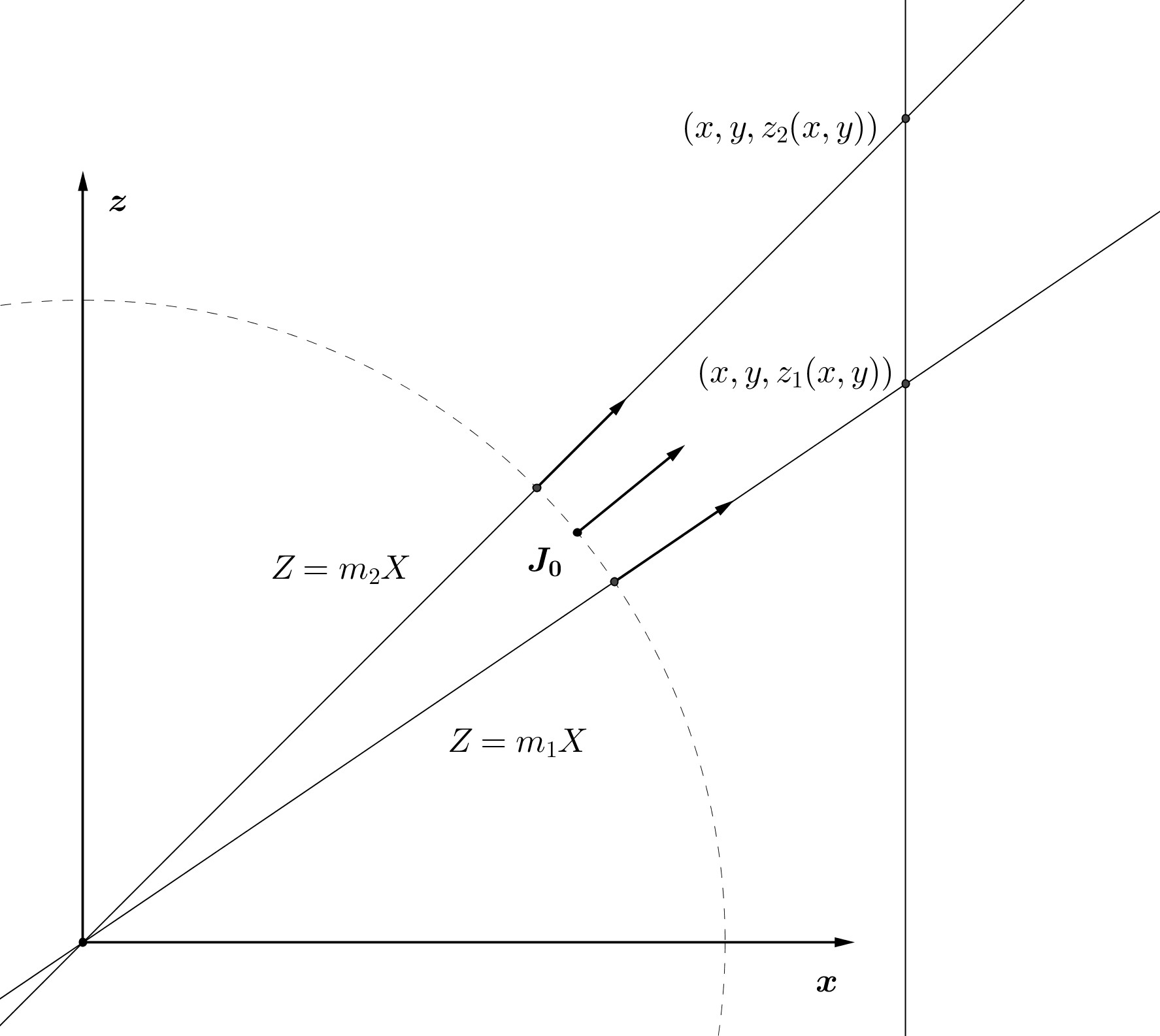}\caption{\label{fig:Flow J0}Flow $\protect\JJ_{0}$ of $P_{0}=\log(X^{2}+Y^{2})$}
\end{figure}

\begin{figure}
\includegraphics{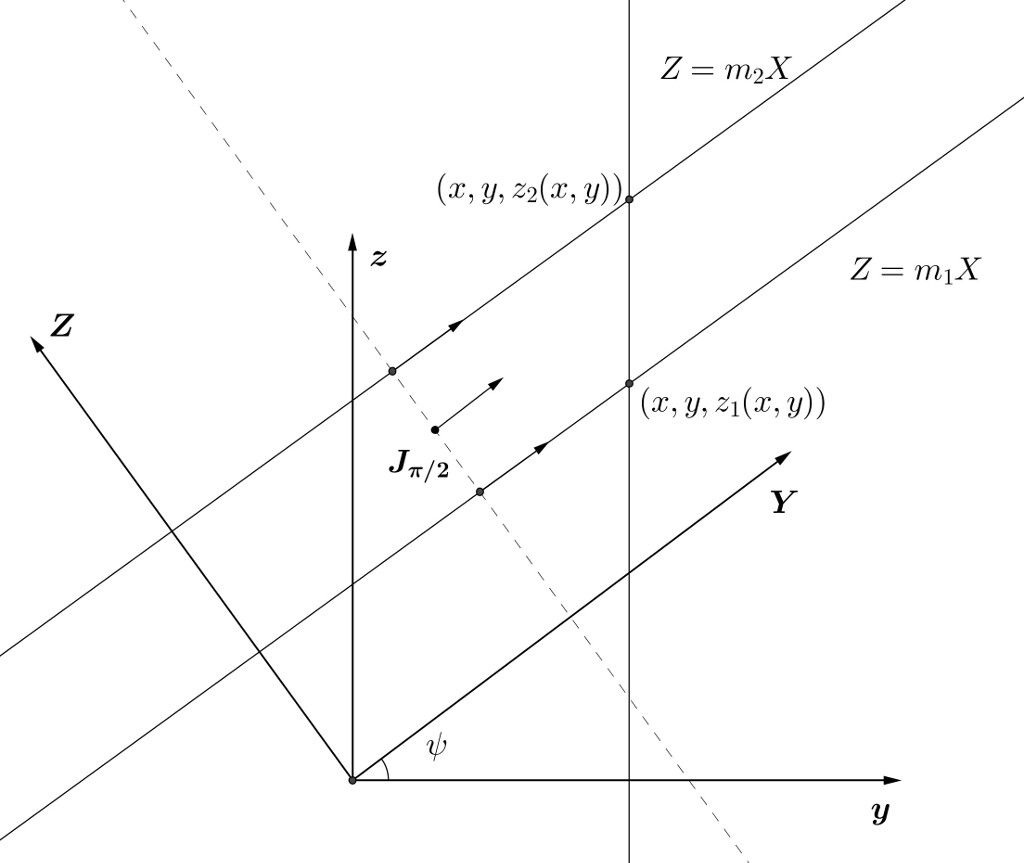}\caption{\label{fig:Flow of Jpi2}Flow $\protect\JJ_{\frac{\pi}{2}}$ of $P_{\frac{\pi}{2}}=Y$. }
\end{figure}

Consider two planes with non-parallel normal vectors $\nn_{1},\nn_{2}$
and let $\zz$ be the unit vector along the direction of projection.
The intersection of the planes is spanned by the unit vector
\[
\nn=\frac{\nn_{1}\times\nn_{2}}{|\nn_{1}\times\nn_{2}|}.
\]
Consider the ortho-normal basis (see Figure \ref{fig:Projection frame})
\begin{equation}
\xx=\frac{\nn\times\zz}{|\nn\times\zz|},\yy=\zz\times\xx\and\zz.\label{eq:littleframe}
\end{equation}
and denote the coordinates is this basis by $x,y$ and $z$. Since
$\nn$ is unitary and orthogonal to $\xx$ we can write
\[
\nn=\cos(\psi)\yy+\sin(\psi)\zz,
\]
where the angle $\psi$, which we will refer to as the \emph{tilt},
is given by 
\begin{equation}
\psi=\arcsin(\nn\cdot\zz)\and-\frac{\pi}{2}\leq\psi\leq\frac{\pi}{2}.\label{eq:Tilt Formula}
\end{equation}
The tilt is simply the angle that the intersection line of the two
planes forms with the projection plane. If $X,Y$ and $Z$ are the
coordinates in the frame 
\begin{eqnarray*}
\XX & = & \xx,\\
\YY=\nn & = & \cos(\psi)\yy+\sin(\psi)\zz,\\
\ZZ & = & -\sin(\psi)\yy+\cos(\psi)\zz.
\end{eqnarray*}
then we have that
\begin{eqnarray*}
X & = & x,\\
Y & = & y\cos(\psi)+z\sin(\psi),\\
Z & = & -y\sin(\psi)+z\cos(\psi).
\end{eqnarray*}
Using these coordinates we can construct a family of functions (parametrized
by $0\leq\omega<2\pi$) as follows 
\[
Q_{\omega}=\cos(\omega)\log(X^{2}+Z^{2})/2+\sin(\omega)Y.
\]
By construction 
\begin{eqnarray*}
Q_{0} & = & \log(X^{2}+Z^{2})\\
Q_{\frac{\pi}{2}} & = & Y
\end{eqnarray*}
are harmonic functions satisfying reflective boundary conditions with
respect to the planes with normal vectors $\nn_{1}$ and $\nn_{2}$
(see Figures \ref{fig:Flow J0} and \ref{fig:Flow of Jpi2}), and
hence so it is $Q_{\omega}$ for all $0\leq\omega<2\pi$.

\subsection*{Computing the effective diffusion operator}

We have that
\begin{eqnarray*}
\der{Q_{\omega}}x & = & \frac{X\cos(\omega)}{x^{2}+Z^{2}}\\
\der{Q_{\omega}}y & = & \cos(\psi)\sin(\omega)-\frac{Z\sin(\psi)\cos(\omega)}{x^{2}+Z^{2}},
\end{eqnarray*}
and hence 

\begin{eqnarray*}
\int_{z_{1}}^{z_{2}}\der{Q_{\omega}}xdz & = & \left[\arctan(Z/x)\sec(\psi)\cos(\omega)\right]_{z=z_{1}}^{z=z_{2}},\\
\int_{z_{1}}^{z_{2}}\der{Q_{\omega}}ydz & = & \left[z\cos(\psi)\sin(\omega)-\frac{1}{2}\log(x^{2}+Z^{2})\tan(\psi)\cos(\omega)\right]_{z=z_{1}}^{z=z_{2}}.
\end{eqnarray*}
The functions $z_{i}$'s are obtained by solving the equation $Z=m_{i}X$
(see Figures \ref{fig:Flow J0} and \ref{fig:Flow of Jpi2}), which
gives us
\[
z_{i}=(m_{i}x+y\sin(\psi))\sec(\psi),
\]
where
\begin{equation}
m_{i}=-\frac{\nn_{i}\cdot\XX}{\nn_{i}\cdot\ZZ}=\frac{\nn_{i}\cdot\xx}{(\nn_{i}\cdot\yy)\sin(\psi)-(\nn_{i}\cdot\zz)\cos(\psi)}\label{eq:Slope Formulas}
\end{equation}
From the above equations we obtain that
\[
\int_{z_{1}}^{z_{2}}\der{Q_{\omega}}xdz=(\arctan(m_{2})-\arctan(m_{1}))\sec(\psi)\cos(\omega)
\]
and
\begin{eqnarray*}
\int_{z_{1}}^{z_{2}}\der{Q_{\omega}}xdz & = & (m_{2}-m_{1})x\sin(\omega)\\
 & - & \log\left(\frac{\sqrt{1+m_{2}^{2}}}{\sqrt{1+m_{1}^{2}}}\right)\tan(\psi)\cos(\omega).
\end{eqnarray*}

If we let
\[
(v_{\omega,1},v_{\omega,2})=\grad\left(\frac{q_{\omega}(x,y)}{w(x,y)}\right)
\]
then
\begin{eqnarray*}
v_{\omega,1} & = & \left(\frac{m_{1}+m_{2}}{2}\right)\sin(\omega)\tan(\psi)+\cos(\omega)/x,\\
v_{\omega,2} & = & \sec(\psi)\sin(\omega).
\end{eqnarray*}
We want to solve (for $\DD)$ the equation
\[
\DD\left(\begin{array}{c}
v_{\omega,1}\\
v_{\omega,2}
\end{array}\right)=\left(\begin{array}{c}
j_{\omega,1}\\
j_{\omega,2}
\end{array}\right).
\]
Since this must hold for all $\omega's$ it must hold in particular
for $\omega=0$ and $\omega=\pi/2$, i.e
\begin{equation}
\DD=\left(\begin{array}{cc}
\DD_{11} & \DD_{12}\\
\DD_{21} & \DD_{22}
\end{array}\right)=\left(\begin{array}{cc}
j_{0,1} & j_{\frac{\pi}{2},1}\\
j_{0,2} & j_{\frac{\pi}{2},2}
\end{array}\right)\left(\begin{array}{cc}
v_{0,1} & v_{\frac{\pi}{2},1}\\
v_{0,2} & v_{\frac{\pi}{2},2}
\end{array}\right)^{-1},\label{eq:Effective Diffusion Coefficient}
\end{equation}
which implies that
\[
\DD=\left(\begin{array}{cc}
\DD_{11} & \DD_{12}\\
\DD_{21} & \DD_{11}
\end{array}\right),
\]
where
\begin{eqnarray*}
\DD_{11} & = & D_{0}\left(\frac{\arctan(m_{2})-\arctan(m_{1})}{m_{2}-m_{1}}\right)\\
\DD_{12} & = & -D_{0}\left(\frac{\arctan(m_{2})-\arctan(m_{1})}{m_{2}-m_{1}}\right)\left(\frac{m_{1}+m_{2}}{2}\right)\sin(\psi),\\
\DD_{21} & = & -D_{0}\left(\frac{\sin(\psi)}{m_{2}-m_{1}}\right)\log\left(\frac{\sqrt{1+m_{2}^{2}}}{\sqrt{1+m_{1}^{2}}}\right),\\
\DD_{22} & = & D_{0}\left(\cos^{2}(\psi)+\frac{1}{2}\left(\frac{m_{1}+m_{2}}{m_{2}-m_{1}}\right)\log\left(\frac{\sqrt{1+m_{2}^{2}}}{\sqrt{1+m_{1}^{2}}}\right)\sin^{2}(\psi)\right).
\end{eqnarray*}
We can write the matrix of $\DD$ more compactly as
\[
\DD=D_{0}\left(\begin{array}{cc}
\omega & -\omega\mu\sin(\psi)\\
-\rho\sin(\psi) & \cos^{2}(\psi)+\mu\rho\sin^{2}(\psi)
\end{array}\right),
\]
where
\[
\rho+i\omega=\left(\frac{1}{m_{2}-m_{1}}\right)\log\left(\frac{1+im_{2}}{1+im_{1}}\right)\and\mu=\frac{1}{2}(m_{1}+m_{2}).
\]

\begin{rem*}
Recall the the complex logarithm function is defined by 
\[
\log(x+iy)=\log(\sqrt{x^{2}+y^{2}})+i\arctan(y/x).
\]

\end{rem*}
The scalars $\DD_{ij}$'s are the coefficients of the linear operator
$\DD$ in the ortho-normal frame formed by $\xx$ and $\yy$, i.e
\begin{eqnarray}
\DD(\xx) & = & \DD_{11}\xx+\DD_{12}\yy,\label{eq:D in Frame 1}\\
\DD(\yy) & = & \DD_{21}\xx+\DD_{22}\yy.\label{eq:D in Frame 2}
\end{eqnarray}

\subsection*{The diffusion ellipsoid and principal response lines}

Consider a square matrix $A$ with real coefficients and $\det(A)\not=0$.
The matrix 
\[
S_{A}=(AA^{T})^{1/2}
\]
is well defined since $AA^{T}$ is symmetric and positive definite.
If we let
\[
R_{A}=S_{A}^{-1}A
\]
then we can write 
\[
A=S_{A}R_{A},
\]
where the matrix $R_{A}$ is orthogonal (since $R_{A}R_{A}^{T}=S_{A}^{-1}AA^{T}S_{A}^{-1}=I$).
In words, any square matrix is the product of a symmetric matrix and
an orthogonal one. 

\begin{figure}

\includegraphics[scale=0.7]{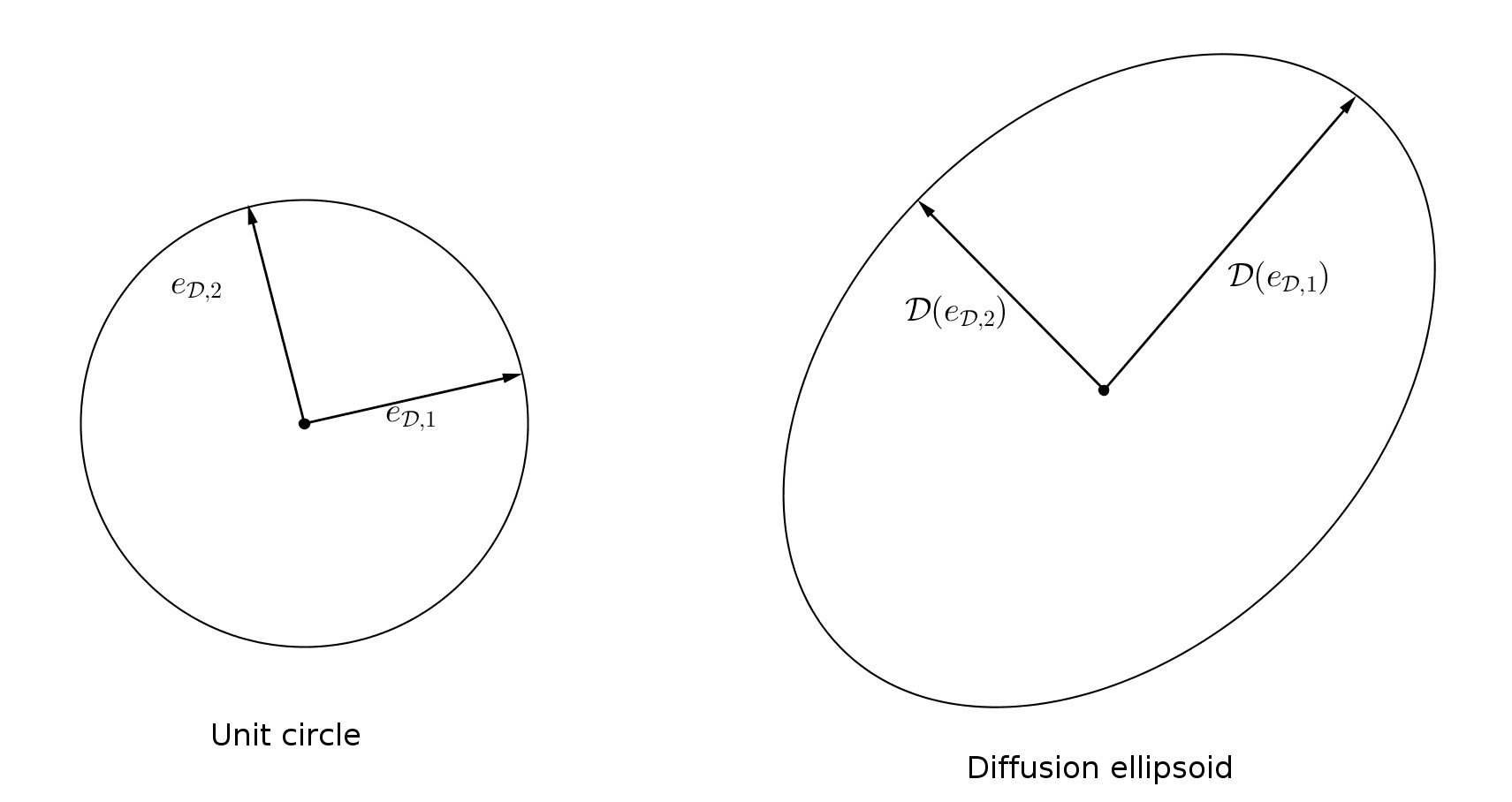}\caption{\label{fig:Principal-response-directions}Principal response directions}

\end{figure}

If we apply the above result to $\DD$ we obtain the decomposition
\[
\DD=S_{\DD}R_{\DD}.
\]
Hence, the circle of unit vectors maps under $\DD$ to an ellipsoid
$E_{\DD}$ having its mayor and minor axes aligned with the eigen-vectors
of $S_{\DD}$ and the corresponding widths are given by the eigenvalues
of $S_{\DD}$. We will refer to $E_{\DD}$ as the \emph{effective
diffusion ellipsoid}. For unit vectors $\ff_{\DD,1}$ and $\ff_{\DD,2}$
aligned with the mayor and minor axes of $E_{\DD}$ the unit vectors
$\ee_{\DD,1}=R_{\DD}^{T}\ff_{1}$ and $\ee_{\DD,2}=R^{T}\ff_{2}$
map under the action of $\DD$ to vectors whose end points lay in
$E_{\DD}$ and aligned with its mayor and minor axes (see Figure \ref{fig:Principal-response-directions}).
We will refer to the lines spanned by $\ee_{\DD,1},\ee_{\DD,2}$ as
\emph{principal response lines} of $\DD$. The physical meaning of
these lines can be obtained by recalling Fick's formula
\[
\jj=-\DD\grad p,
\]
so that when $\grad p$ is aligned with $\ee_{\DD,1}$ the corresponding
relative flux magnitude $|\jj|/|\grad p|$ is maximal. Similarly,
when $\grad p$ is aligned with $\ee_{\DD,2}$ the value of $|\jj|/|\grad p|$
is minimal. 

\subsection*{The case of parallel planes}

In this case $\nn_{1}=\nn_{2}$, which implies that $m_{1}=m_{2}=\mu$.
In particular, the vector 
\[
\nn=\frac{\nn_{1}\times\nn_{2}}{|\nn_{1}\times\nn_{2}|}
\]
is not well defined and can not used to construct the frame $\xx,\yy$
(see \ref{eq:littleframe}). Instead, for
\[
\nn_{1}=\nn_{2}=(a,b,c)
\]
we let 
\[
\nn=\begin{cases}
\frac{1}{\sqrt{a^{2}+b^{2}}}(-b,a,0) & \hbox{if\,\,\,\,}a^{2}+b^{2}>0,\\
(0,1,0) & \hbox{if\,\,\,\,}a^{2}+b^{2}=0.
\end{cases}
\]
so that for $a^{2}+b^{2}>0$ we get
\[
\xx=\frac{1}{\sqrt{a^{2}+b^{2}}}(a,b,0),\yy=\frac{1}{\sqrt{a^{2}+b^{2}}}(-b,a,0)
\]
and for $a^{2}+b^{2}=0$ we get
\[
\xx=(1,0,0)\and\yy=(0,1,0).
\]
In both cases we have that $\psi=0$ (see \ref{eq:Tilt Formula})
and since

\begin{eqnarray*}
\rho+i\omega & = & \lim_{m_{1},m_{2}\mapsto\mu}\left(\frac{1}{m_{2}-m_{1}}\right)\log\left(\frac{1+im_{2}}{1+im_{1}}\right),\\
 & = & \frac{d}{d\mu}(\log(1+i\mu))=\frac{i}{1+i\mu}\\
 & = & \frac{\mu+i}{1+\mu^{2}},
\end{eqnarray*}
we obtain
\begin{equation}
\DD=D_{0}\left(\begin{array}{cc}
\frac{1}{1+\mu^{2}} & 0\\
0 & 1
\end{array}\right),\label{eq:ParallelPlanesD}
\end{equation}
where $\mu$ is the common slope of the two planes with respect to
the $x,y$-plane. When $\mu=0$ the planes are parallel to the $x,y$-plane
and the effective diffusion operator is simply scalar multiplication
by $D_{0}$, i.e the diffusion equation remains unchanged after the
projection procedure. This is to be expected, as in this case the
walls have no effect on the diffusion process in the main direction
of transport.

\subsection*{The case of non-parallel planes with no tilt}

\begin{figure}
\includegraphics{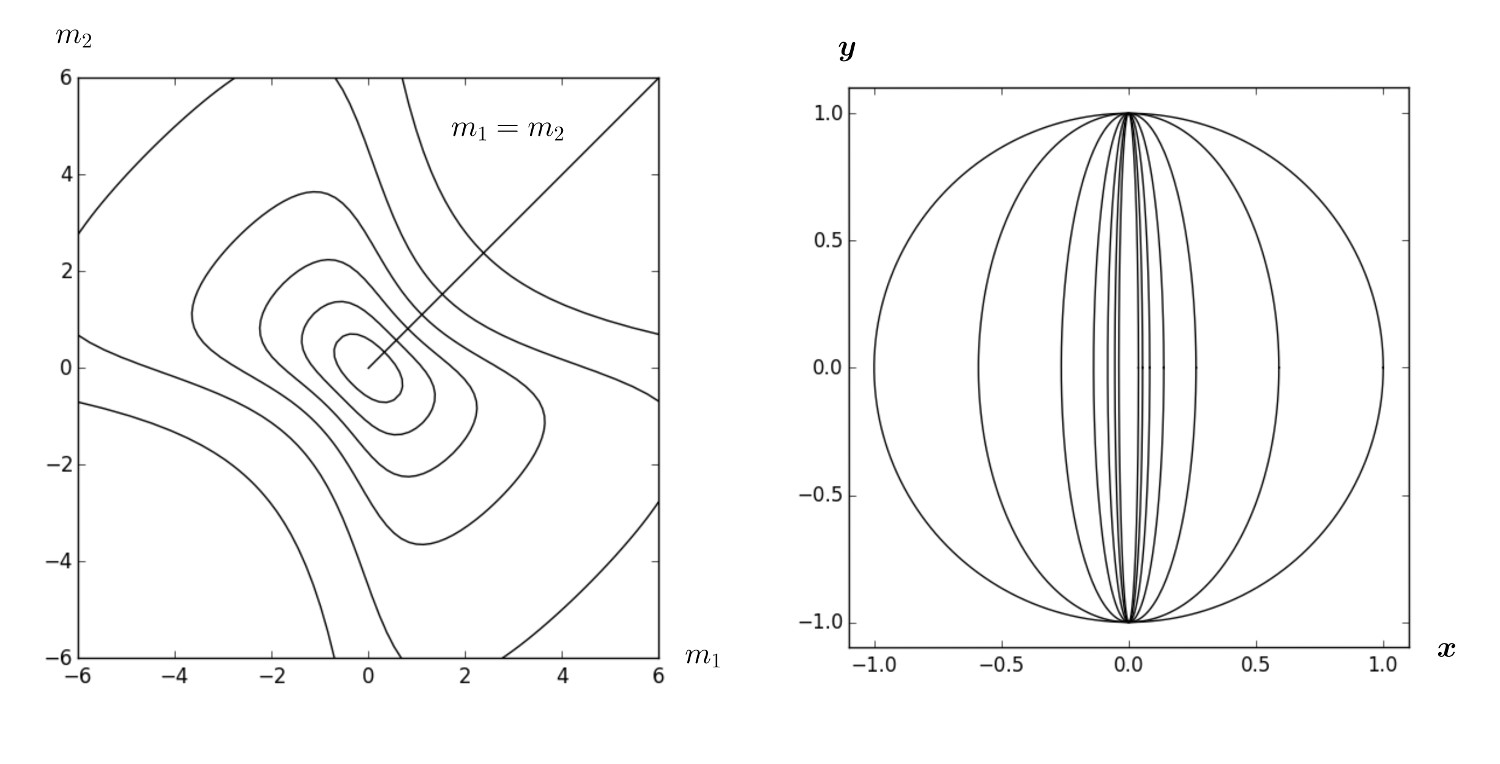}\caption{\label{fig:Effective-diffusion-ellipsoids psi=00003D0}Effective diffusion
ellipsoids for $\psi=0$ and $D_{0}=1$. Each level curve of $\omega$
(left) corresponds to a single ellipsoid (right). As we move along
the ray $m_{1}=m_{2}=\mu\geq0$ we obtain ellipsoids with semi-axis
$1/(1+\mu^{2})$ and $1$.}
\end{figure}
In this case we have $m_{1}\not=m_{2}$ and $\psi=0$, i.e the two
planes are non-parallel and their intersection line is parallel to
the $x,y$-plane. We then have that
\begin{equation}
\DD=D_{0}\left(\begin{array}{cc}
\omega & 0\\
0 & 1
\end{array}\right),\label{eq:NoTiltD}
\end{equation}
where
\[
\omega=\frac{\arctan(m_{2})-\arctan(m_{1})}{m_{2}-m_{1}}.
\]
Observe that when $m_{1}=m_{2}$ formula \ref{eq:NoTiltD} becomes
\ref{eq:ParallelPlanesD}.

The effective diffusion ellipsoid $E_{\DD}$ has $\xx$ and $\yy$
as principal axes and corresponding widths $D_{0}\omega$ and $D_{0}$
. The dependence of $E_{\DD}$ on $m_{1}$ and $m_{2}$ is shown in
Figure \ref{fig:Effective-diffusion-ellipsoids psi=00003D0}.

\subsection*{The case of extreme tilts}

This occurs when $\psi=-\pi/2$ or $\psi=\pi/2$, i.e the planes are
orthogonal to the $x,y$-plane, and we have corresponding diffusion
matrices
\[
\DD_{-}=D_{0}\left(\begin{array}{cc}
\omega & \mu\omega\\
\rho & \mu\rho
\end{array}\right)\hbox{\,\,\,or\,\,\,}\DD_{+}=D_{0}\left(\begin{array}{cc}
\omega & -\mu\omega\\
-\rho & \mu\rho
\end{array}\right).
\]
Both matrices share the same eigen-values, given by
\[
0,D_{0}(\mu\rho+\omega).
\]
The eigen-vectors of $\DD_{-}$are 
\[
(\mu,-1)\and(\text{\ensuremath{\omega},}\rho),
\]
and those of $\DD_{+}$ are
\[
(\mu,1)\and(\omega,-\rho).
\]
If $E_{-}=E_{\DD_{-}}$ and $E_{+}=E_{\DD_{+}}$ then $E_{-}$ is
a line segment (a degenerate ellipsoid) joining the diametrically
opposite points
\[
\pm D_{0}\left(\frac{\mu\rho+\omega}{(\omega^{2}+\rho^{2})^{1/2}}\right)(\omega,\rho)
\]
and $E_{+}$ is a line segment joining the diametrically opposite
points
\[
\pm D_{0}\left(\frac{\mu\rho+\omega}{(\omega^{2}+\rho^{2})^{1/2}}\right)(\omega,-\rho).
\]

\subsection*{The case of varying tilt}

\begin{figure}

\includegraphics[scale=0.8]{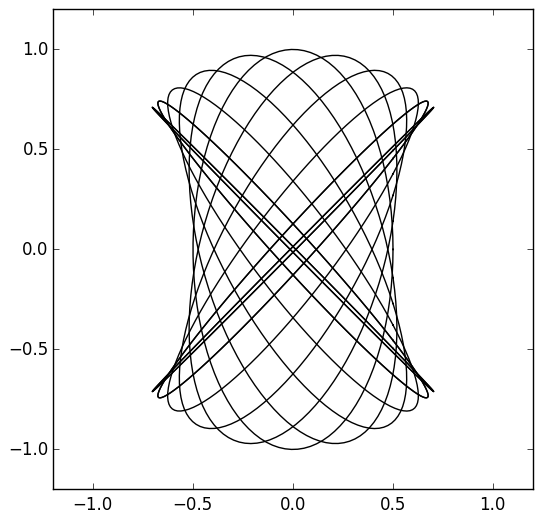}\caption{\label{fig:varyingtilt}Effective diffusion ellipsoids obtained by
varying the tilt parameter}

\end{figure}

If we vary the tilt parameter $-\pi/2<\psi<\pi/2$ the effective diffusion
ellipsoid gives us a family of ellipsoids which in the limiting cases
$\psi=-\pi/2,\pi/2$ become line segments (see Figure \ref{fig:varyingtilt})

\section{The effective diffusion matrix for two surfaces}

For surfaces defined by $z=z_{i}(x,y)$ we chose their unit normal
vectors as 
\[
\nn_{1}=-\frac{(-\grad z_{1},1)}{\sqrt{1+|\grad z_{1}|^{2}}}\and\nn_{2}=\frac{(-\grad z_{2},1)}{\sqrt{1+|\grad z_{2}|^{2}}},
\]
where the signs are so that $\nn_{1}$ and $\nn_{2}$ point the outside
of the region defined by the formula $z_{1}(x,y)\leq z_{2}(x,y).$
We then have that
\[
\nn=\nn_{1}\times\nn_{2}=\frac{\left(\grad^{\bot}w,\grad z_{1}\cdot\grad^{\bot}z_{2}\right)}{\sqrt{1+|\grad z_{1}|^{2}}\sqrt{1+|\grad z_{2}|^{2}}},
\]
where the width function $w$ is given by 
\[
w=z_{2}-z_{1},
\]
and the operators $\grad$ and $\grad^{\bot}$ are given by 
\[
\grad=\left(\dero x,\dero y\right)\and\grad^{\bot}=\left(-\dero y,\dero x\right).
\]
We will refer to $\grad^{\bot}$ as the \emph{orthogonal gradient
operator} (observe that $\grad^{\bot}w$ is the Hamiltonian vector
field of $w$). For $\zz=(0,0,1)$ we have that
\[
\xx=\frac{\nn\times\zz}{|\nn\times\zz|}=\left(\frac{\grad w}{|\grad w|},0\right)\and\yy=\zz\times\xx=\left(\frac{\grad^{\bot}w}{|\grad w|},0\right).
\]
Since $\xx$ and $\yy$ are in the plane $z=0$ we will simply write
\begin{equation}
\xx=\frac{\grad w}{|\grad w|}\and\yy=\frac{\grad^{\bot}w}{|\grad^{\bot}w|}.\label{eq:XYFit}
\end{equation}
Formulas \ref{eq:D in Frame 1} and \ref{eq:D in Frame 2} imply that
\begin{eqnarray*}
\DD(\grad w) & = & \DD_{11}\grad w+\DD_{12}\grad^{\bot}w,\\
\DD(\grad^{\bot}w) & = & \DD_{21}\grad w+\DD_{22}\grad^{\bot}w.
\end{eqnarray*}

We now compute $m_{1},m_{2}$ and $\psi$, since we need them to compute
the coefficients $\DD_{ij}$'s . Using the formulas \ref{eq:XYFit}
, \ref{eq:Tilt Formula} and \ref{eq:Slope Formulas} we obtain
\begin{equation}
\psi=\arcsin\left(\frac{\grad z_{1}\cdot\grad^{\bot}z_{2}}{\sqrt{1+|\grad z_{1}|^{2}}\sqrt{1+|\grad z_{2}|^{2}}}\right).\label{eq:Fitted Tilt}
\end{equation}
and
\begin{equation}
m_{i}=\frac{\grad z_{i}\cdot\grad w}{(\grad z_{i}\cdot\grad^{\bot}w)\sin(\psi)+|\grad w|\cos(\psi)}.\label{eq:FittedSlopes}
\end{equation}

\section{examples}

\subsection*{Recovering the 2-dimensional case}

We assume that $z_{1}$ and $z_{2}$ are only functions of the $x$-variable
and write $z_{i}=z_{i}(x)$, so that
\[
\grad z_{i}=(z'_{i},0).
\]
This implies that the vectors $\grad z_{1},\grad z_{2}$ and $\grad w$
are all parallel, that the tilt $\psi$ is identically zero and that
\[
m_{i}=\frac{z'_{i}(z'_{2}-z'_{1})}{|z'_{2}-z'_{1}|}=\pm z'_{i},
\]
where the plus sign is chosen at the points $(x,y)$ where $z'_{2}>z'_{1}$
(i.e $w'>0)$ and the minus sign at the points $(x,y)$ where $z'_{2}<z'_{1}$
(i.e $w'<0$). We conclude that (irrespective of the sign in the above
formula) $\DD$ is given by
\begin{equation}
\DD=D_{0}\left(\begin{array}{cc}
\frac{\arctan(z'_{2})-\arctan(z'_{2})}{z'_{2}-z'_{1}} & 0\\
0 & 1
\end{array}\right).\label{eq:chanelrecovery}
\end{equation}
The above formula for $\DD$ is its representations in basis formed
by $\grad w$ and $\grad^{\bot}w$, but it is easy to see that in
this case the same matrix represents $\DD$ in the basis formed by
the vectors $(1,0)$ and $(0,1)$. Hence, we have obtained the result
in \cite{kn:di-projection-diffusion} as a particular case of our
formula for surfaces. 
\begin{rem*}
At the points $(x,y)$ where $z'_{2}=z'_{1}$ (i.e $w'=0)$ we have
that
\[
\DD=\left(\begin{array}{cc}
\frac{1}{1+(z'_{1})^{2}} & 0\\
0 & 1
\end{array}\right)
\]

\end{rem*}

\subsection*{Surfaces with vanishing tilt function}

The tilt function $\psi$ vanishes if $\grad z_{1}$ and $\grad z_{2}$
are parallel vectors at all points. This holds if we can write 
\[
z_{i}(x,y)=f_{i}(z(x,y)),
\]
and in this case we have that
\[
\grad z_{i}=f'_{i}\grad z\and\grad w=(f'_{2}-f'_{1})\grad z.
\]
Using formula \ref{eq:FittedSlopes} and assuming $f'_{1}\not=f_{2}'$
we obtain 
\begin{equation}
m_{i}=\pm f'_{i}|\grad z|,\label{eq:slopesforzerotilt}
\end{equation}
where the plus sign is selected if $f'_{2}<f'_{1}$ and the minus
sign if $f'_{2}>f'_{1}$. We can then write
\[
\DD=D_{0}\left(\begin{array}{cc}
\text{\ensuremath{\omega}} & 0\\
0 & 1
\end{array}\right)
\]
where (and independently of the choice of sign in formula \ref{eq:slopesforzerotilt})
we have that
\[
\omega=\frac{\arctan(f'_{2}|\grad z|)-\arctan(f'_{1}|\grad z|)}{(f'_{2}-f'_{1})|\grad z|},
\]
at the points $(x,y)$ at which$f'_{1}\not=f'_{2}$, and 
\[
\omega=\frac{1}{1+(f_{1}'|\grad z|)^{2}}=\frac{1}{1+(f_{2}'|\grad z|)^{2}}
\]
at the points $(x,y)$ where $f'_{1}=f'_{2}$. 

\begin{figure}
\includegraphics[scale=0.45]{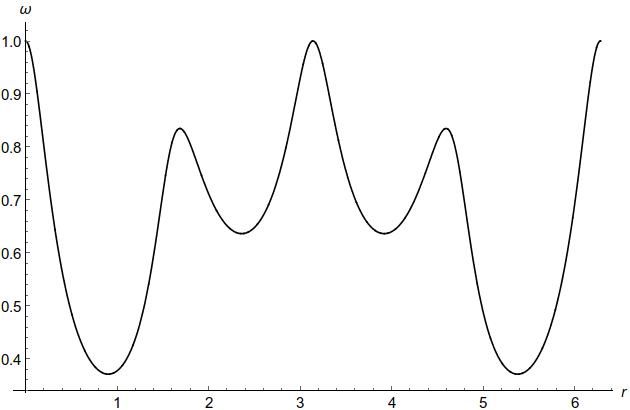}\caption{\label{fig:RadialDiffusionExample}Radial effective diffusion function
of the surfaces $z_{1}=\sin(\sqrt{x^{2}+y^{2}})-3/2\protect\and z_{2}=\cos(2\sqrt{x^{2}+y^{2}})+3/2$. }

\end{figure}

For example, if we let
\[
z_{1}=\sin(\sqrt{x^{2}+y^{2}})-3/2\and z_{2}=\cos(2\sqrt{x^{2}+y^{2}})+3/2,
\]
then 
\[
f_{1}=\sin(r)-3/2,f_{2}=\cos(2r)+3/2\and r=\sqrt{x^{2}+y^{2}}.
\]
In this case the integral curves of $\grad w$ generate rays of the
form
\[
r\mapsto r(\cos(\text{\ensuremath{\theta}),\ensuremath{\sin}(\ensuremath{\theta}))}\where r\geq0
\]
and those of $\grad^{\bot}w$ generate circles of the form
\[
\theta\mapsto r(\cos(\theta),\sin(\theta))\where0\leq\theta<2\pi.
\]
Along the directions of the circles the diffusion process has effective
diffusion constant equal to $D_{0}$, and along the direction defined
by the rays the diffusion process has effective diffusion function
$D_{0}\omega$ (see Figure \ref{fig:RadialDiffusionExample}) .
\begin{rem*}
Observe that in this case the matrix $\DD$ is diagonal. Hence the
principal response lines are generated by the vector fields $\grad^{\bot}w$
and $\grad w$.
\end{rem*}

\subsection*{An example with non-vanishing tilt function}

\begin{figure}
\includegraphics[scale=0.23]{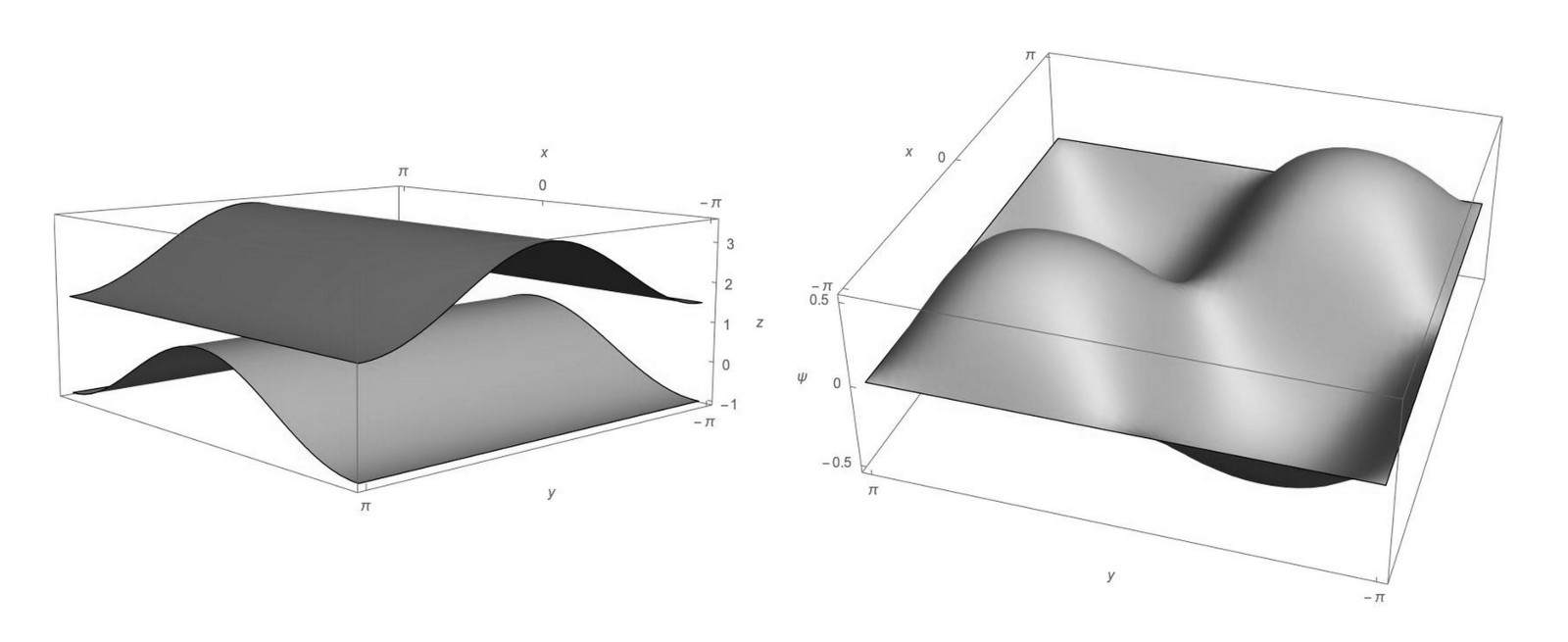}\caption{\label{fig:WavesAndTilt}Two orthogonal planar waves (left) and their
tilt function (right)}
\end{figure}

Consider the orthogonal planar wave surfaces given by the functions
\[
z_{1}=\cos(x)\and z_{2}=\cos(y)+5/2.
\]
The tilt function is given by (see Figure \ref{fig:WavesAndTilt})
\[
\psi=-\arcsin\left(\frac{\sin(x)\sin(y)}{\sqrt{(1+\sin(x)^{2})(1+\sin(y)^{2})}}\right),
\]
which only vanishes on the lines $x=n\pi$ and $y=m\pi$ for any integers
$m$ and $n$.

\begin{figure}
\includegraphics[scale=0.23]{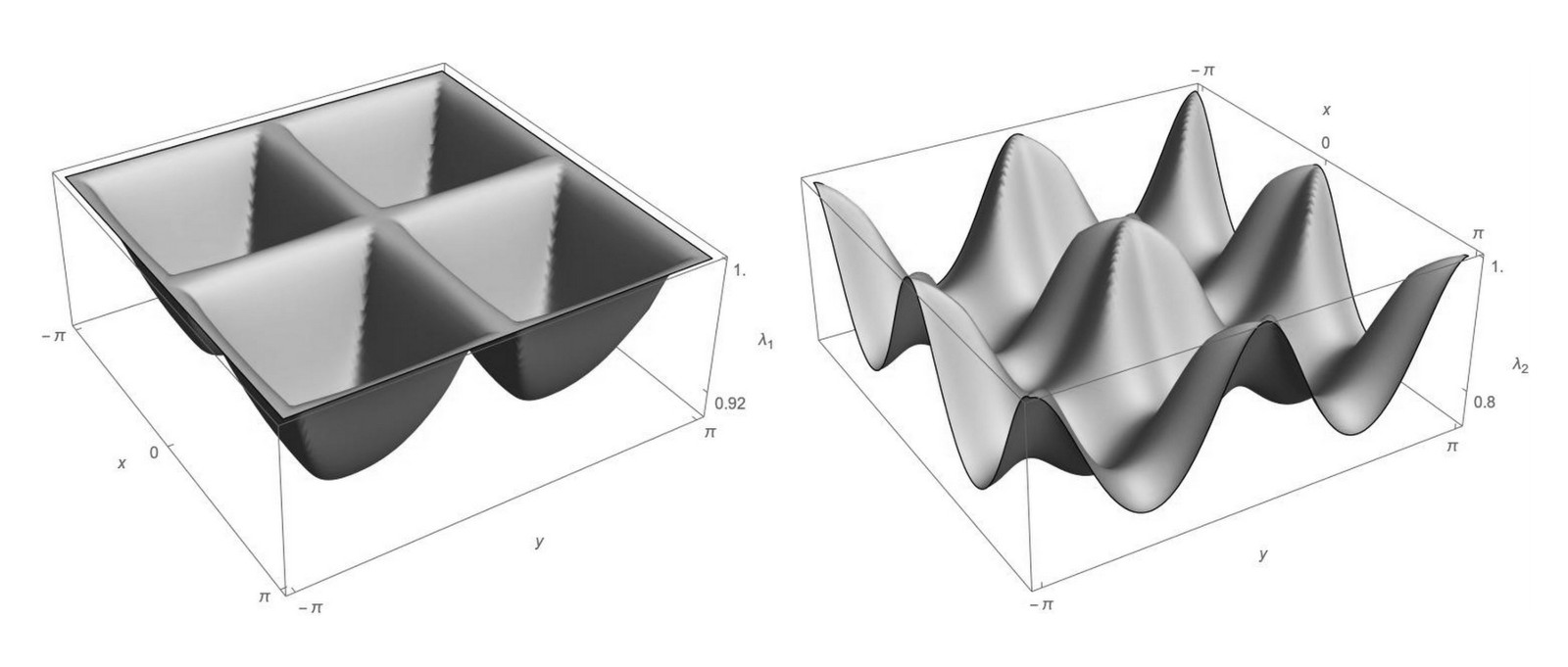}\caption{\label{fig:WavesEigenvalues}Eigenvalue functions of two planar orthogonal
waves}
\end{figure}

\begin{figure}

\includegraphics[scale=0.23]{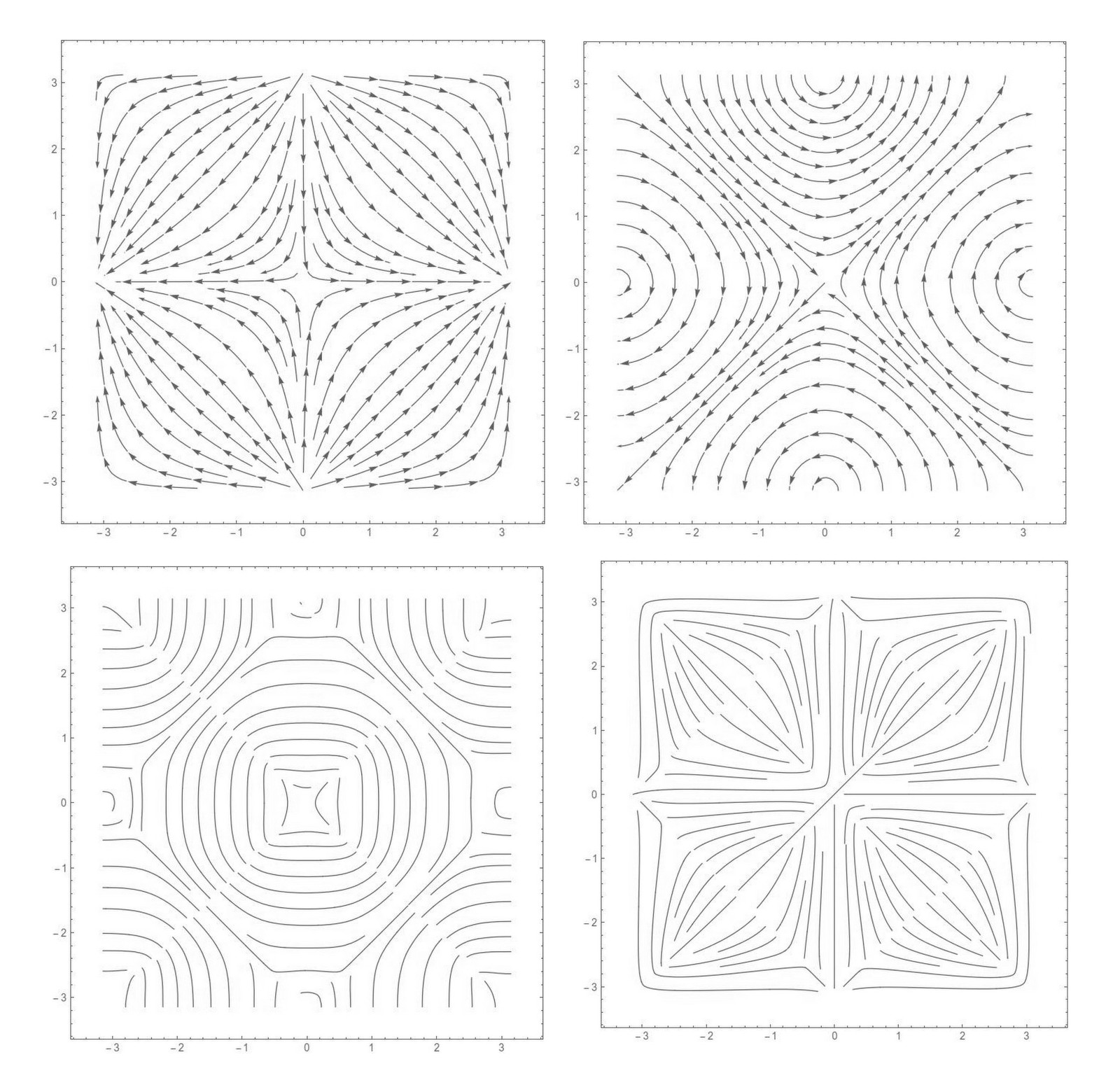}\caption{\label{fig:WavesFields} Vector and line fields of two orthogonal
wave-like functions. The vector fields $\protect\grad w$ (upper left
corner) and $\protect\grad^{\bot}w$ (upper right corner), and the
line fields spanned by $\protect\ee_{\protect\DD,1}$ (lower left
corner) and $\protect\ee_{\protect\DD,2}$ (lower right corner).}

\end{figure}

Recall that we have a decomposition of $\DD$ of the form
\[
\DD=S_{\DD}R_{\DD}
\]
where $S_{\DD}$ is symmetric and $R_{\DD}$ is orthogonal. The eigenvalue
functions $\lambda_{1}$ and $\lambda_{2}$ of $S_{\DD}$ are shown
in figure \ref{fig:WavesEigenvalues}. In Figure \ref{fig:WavesFields}
we show the gradient fields $\grad w$ and $\grad^{\bot}w$ of the
width function $w=z_{2}-z_{1}$, and the fields of principal directions
of $\DD$. In this case $\grad^{\bot}w$ and $\grad w$ do not in
general span the principal responses lines, since the tilt function
does not vanishes at all points.

\section{Conclusions and Future work}

We have obtained the effective diffusion matrix which results of projecting
a diffusion process between two surfaces (of the form $z=z_{1}(x,y)$
and $z_{2}(x,y)$) onto the $x,y$-plane. In future work we plan to
project the diffusion process into the middle surface $z_{1/2}=(z_{1}+z_{2})/2$,
where we expect to obtain a more clear understanding of the effective
diffusion operator. The added complication is that the differential
geometry of the middle surface is now non-trivial, in contrast to
the $x,y$-plane which is a flat.

\section{Aknowledgments}

This work was supported by PROMEP Grant No. UGTP-PTC-498, and CONACYT
Grants No. 135106 and No. 222870.

\bibliographystyle{plain}
\bibliography{myBib}

\begin{thebibliography}{1}

\bibitem{kn:bradley}
R.M. Bradley.
\newblock Diffusion in a two-dimensional channel with curved midline and
  varying width.
\newblock {\em Phys. Rev. E}, B 80, 2009.

\bibitem{va:fj3dcurves}
C.Valero and R.Herrera.
\newblock Fick-jacobs equation for channels over three-dimensional curves.
\newblock {\em Phy}, 90(052141), 2014.

\bibitem{kn:di-projection-diffusion}
L.~Dagdug and I.~Pineda.
\newblock Projection of two-dimensional diffusion in a curved midline and
  narrow varying width channel onto the longitudinal dimension.
\newblock {\em The Journal of Chemical Physics}, 137, 2012.

\bibitem{kn:kp-diffusion-projection}
P.~Kalinay and K.~Percus.
\newblock Projection of a two-dimensional diffusion in a narrow channel onto
  the longitudinal dimension.
\newblock {\em The Journal of Chemical Physics}, 122, 2005.

\bibitem{kn:aproximations}
P.~Kalinay and K.~Percus.
\newblock Aproximations to the generalized fick-jacobs equation.
\newblock {\em Physical Review E}, 78, 2008.

\bibitem{kn:ogawa}
N.~Ogawa.
\newblock Diffusion in a curved cube.
\newblock {\em Physics Letters A}, 377:2465--2471, 2013.

\bibitem{kn:reguerarubi}
D.~Reguera and J.M. Rubí.
\newblock Kinetic equations for diffusion in the prescence of entropic
  barriers.
\newblock {\em Physical Review E}, 64, 2001.

\bibitem{va:edfibrebundles}
C.~Valero.
\newblock Effective diffusion on riemannian fiber bundles.
\newblock {\em J. Math. Phys}, 56(023507), 2015.

\bibitem{kn:entropybarrierzwanzig}
Robert Zwanzig.
\newblock Diffusion past an entropy barrier.
\newblock {\em J. Phys. Chem.}, 96:3926--3930, 1992.

\end{thebibliography}

\end{document}